# Evidence for a Superglass State in Solid $^4$He

B. Hunt,[1§] E. Pratt,[1§] V. Gadagkar,[1] M. Yamashita,[1,2] A. V. Balatsky[3] and J.C. Davis[1,4]

[1] *Laboratory of Atomic and Solid State Physics, Department of Physics, Cornell University, Ithaca, NY 14853, USA.*

[2] *Department of Physics, Kyoto University, Kyoto 606-8502, Japan.*

[3] *T-Division, Center for Integrated Nanotechnologies, MS B 262, Los Alamos National Laboratory, Los Alamos, NM 87545, USA.*

[4] *Scottish Universities Physics Alliance, School of Physics and Astronomy, University of St. Andrews, St. Andrews, Fife KY16 9SS, Scotland, UK.*

[§]These authors contributed equally to this work.

**Although solid helium-4 ($^4$He) may be a supersolid it also exhibits many phenomena unexpected in that context. We studied relaxation dynamics in the resonance frequency *f*(*T*) and dissipation *D*(*T*) of a torsional oscillator containing solid $^4$He. With the appearance of the "supersolid" state, the relaxation times within *f*(*T*) and *D*(*T*) began to increase rapidly together. More importantly, the relaxation processes in both *D*(*T*) and a component of *f*(*T*) exhibited a complex synchronized ultra-slow evolution towards equilibrium. Analysis using a generalized rotational susceptibility revealed that, while exhibiting these apparently glassy dynamics, the phenomena were quantitatively inconsistent with a simple excitation freeze-out transition because the variation in *f* was far too large. One possibility is that amorphous solid $^4$He represents a new form of supersolid in which dynamical excitations within the solid control the superfluid phase stiffness.**

A 'classic' supersolid (*1-5*) is a bosonic crystal with an interpenetrating superfluid component. Solid $^4$He has long been the focus of searches for this state (*6*). To demonstrate its existence unambiguously, macroscopic quantum phenomena (*7*) such as persistent mass currents, circulation quantization, quantized vortices, or the superfluid Josephson effect must be observed. None of these effects have been detected in solid $^4$He.

There are, however, indications that this material could be a supersolid. This is because high-Q torsional oscillators (TOs) containing solid $^4$He exhibit an increase in resonance frequency $f(T)$ at low temperatures. This is detectable below a temperature $T_C$~65 mK in the purest, most crystalline samples; below $T_C$~300 mK in more amorphous samples; and below at least $T_C$~500 mK when dilute concentrations of $^3$He exist (*8-11*). A strong dissipation peak in $D(T) \equiv Q^{-1}(T)$ occurs in association with the rapid rise of $f(T)$ (*9,10,12*) but their relation has not been explained. These results [now widely reproduced (*12-15*)] can be interpreted as a $^4$He supersolid whose rotational inertia is reduced due to its superfluid component. Support for this interpretation comes from the reductions in the net frequency increase when the TO annuli containing solid $^4$He are blocked (*9, 16*).

But many phenomena inexplicable in the context of a classic superfluid are also observed in equivalent samples of solid $^4$He. These include, for example, maximum dc mass flow rates inconsistent with the TO dynamics (*17-19*), large increases in $T_C$ with the introduction of dilute $^3$He concentrations (*8,11*), strong effects of annealing on the magnitude of TO frequency shifts (*10,12,20*), velocity hysteresis in the frequency shift (*13*), and shear stiffening of the solid coincident with the TO frequency increase (*21*). These phenomena indicate some unanticipated interplay between dynamical degrees of freedom of the solid and any superfluid component.

In response, new theories have been proposed that solid $^4$He is (i) a non-superfluid glass (*22,23*), (ii) a fluid of fluctuating quantum vortices (*24*), (iii) a superfluid network at linked grain boundaries (*25*), (iv) a viscoelastic solid (*26*), or a superglass - a type of granular superfluid within an amorphous solid (*27-30*). To help discriminate between such ideas, we focus on the relaxation dynamics of solid $^4$He, which should distinguish a simple superfluid state from a purely glassy state.

Using a TO containing solid $^4$He (Fig. 1) we measure $f(T)$ and $D(T)$ effects that are in good agreement with those of other research groups (*8-16*). Fig. 1 shows the evolution of $f(T)$ (blue circles) and $D(T)$ (red triangles) for our typical sample; the change in $f(T)$ between 300 and 10 mK would represent a "supersolid fraction" of 4.8% if the frequency shift were entirely attributable to a superfluid decoupling. Our samples, while formed by the "blocked capillary" procedure and therefore amorphous, are of the type most widely studied in the field (*8,9,10,11,13,14,15*). Thus they are representative of the full spectrum of solid $^4$He effects to be explained. Equivalent effects were detected in four different samples studied in two different cells of this type (*31*). All our experiments were performed at a maximum wall velocity of less than 4.5 µm/s.

To examine the relaxational characteristics of solid $^4$He, we perform the experiments outlined in fig. S1 (*31*). The temperature is decreased stepwise from an initial temperature $T_i$ to a final equilibrium temperature $T_{eq}$, and the rapid coevolution of $f$ and $D$ is observed as the thermometers approach $T_{eq}$. More importantly, the subsequent changes after the thermometers equilibrate, $f(t,T_{eq})$ and $D(t,T_{eq})$, are measured. Data from five representative experiments are shown in Fig. 2, A and B, with each trace offset by 5000 seconds for clarity. In Fig. 2A the vertical axis represents the percentage of the total frequency change during each experiment. In Fig. 2B it represents the percentage of the equivalent total dissipation change. The green circles denote the time $t_{eq}$ at which the mixing chamber temperature equilibrates. Though for the initial $t<t_{eq}$ part of each trace both $f(t)$ and $D(t)$ change rapidly with temperature, their slopes change sharply at $t_{eq}$ indicating that the solid inside the TO maintains thermal equilibrium with the mixing chamber.

Before the "supersolid" signature appears, Fig. 2, A and B, reveal that these relaxation rates are independent of temperature and less than 100 s. But below this temperature they begin to increase rapidly. The time constants for relaxation processes in $f$ and $D$, $\tau_f(T)$ and $\tau_D(T)$, are indicated schematically in Fig. 2, A and B. They are measured by fitting the exponential $f(t) = C_1 - C_2 \exp(-t/\tau_f(T))$ and $D(t) = C_3 - C_4 \exp(-t/\tau_D(T))$ to each trace for times $t > t_{eq} = 0$. Both $\tau_f(T)$ and $\tau_D(T)$ increase rapidly on indistinguishable trajectories (Fig. 2C), indicating that the ultraslow relaxation processes in $f$ and $D$ are intimately linked. Such ultra-slow dynamics in the 'supersolid' state have also been observed elsewhere (*20,32,33*). It is difficult to reconcile any of these effects with thermal relaxation in a superfluid. Therefore a better understanding of the relaxation dynamics of amorphous solid $^4$He is required.

We first examine the relation between the relaxation dynamics of dissipation and the frequency shift as both approach their long-time equilibrium states. In the relevant experiment [fig. S2 (*31*)], the $^4$He sample is cooled to 17 mK and then equilibrated for a time $t>20,000$ s to achieve an unchanging state. It is then heated abruptly to a temperature $T$ and the subsequent relaxation dynamics in both $f(t,T)$ and $D(t,T)$ are monitored. The resulting time dependence of dissipation $D(t,T)$ is shown in Fig. 3A. At short times after temperature stabilization, the dissipation increases slightly (dark blue in Fig. 3A). However, these dissipative processes are actually very far out of equilibrium. As time passes, the dissipation slowly increases on a trajectory indicated by the transition from the blue line representing $D(t,T)$ at $t$~50 s to the dark red line representing $D(t,T)$ at ~5000 s. In the same experiment, the time dependence of $f(t,T)$ is also measured (Fig. 3B). It differs from that of $D(t,T)$; at shortest times after stabilization at $T$ the frequency has already changed greatly from its lowest-temperature value (Fig. 3B). This means that much of the frequency change responds immediately to the mixing-chamber temperature change (and therefore also that rapid thermal equilibrium always exists between the sample and the mixing-chamber thermometer). The subsequent evolution of the remaining component of the frequency shift exhibits an ultraslow reduction in $f$ as indicated by the transition from the blue line at $t$~50 s to the dark red line representing t~5000 s in Fig. 3B. These data illustrate how the slowing relaxation dynamics within $D(t,T)$ and $f(t,T)$ are synchronized in such samples of solid $^4$He.

They also imply that thermal hysteresis should occur when temperatures are swept faster than the relevant time constants in Fig. 3, A and B. In Fig. 3C swept-temperature measurements on the same sample show that thermal hysteresis occurs in both $f(T)$ and $D(T)$, with their long-time equilibrium values (solid circles) falling within the hysteresis loops as expected.

These extraordinary relaxation dynamics in $D(t,T)$ and $f(t,T)$ are unexpected in the context of a familiar superfluid. But effects analogous to these are seen during the freeze-out of excitations at a dielectric glass transition (*34*).Thus, the phenomenology of solid $^4$He might also be due to a freeze-out of an ensemble of excitations within the solid (*22*). Indeed, there have been numerous proposals (*27,28,29,30*) that solid $^4$He is a "superglass" – some form of granular superfluid within an amorphous solid.

To examine such hypotheses, we consider the total rotational susceptibility $\chi(\omega,T)$ of the TO plus solid $^4$He sample (22). A classic Debye susceptibility describing the freeze-out of an ensemble of identical excitations with decay time $\tau(T)$ is $\chi_D^{-1}(\omega,T) = g_0/(1-i\omega\tau(T))$. For solid $^4$He, $g_0/\omega_0^2$ would represent the rotational inertia associated with the relevant excitations. Their "back action" on the TO would appear in the total susceptibility as

$$\chi^{-1}(\omega,T) = K - I\omega^2 - i\gamma\omega - g_0/(1-i\omega\tau(T)) \quad (1)$$

where $\gamma$ is the intrinsic damping constant of the TO ($\omega_0 = \sqrt{K/I} = 2\pi f_0$; Fig. 1). The effect of changing the temperature can be captured entirely by the Debye term $\chi_D^{-1}$, whose real and imaginary parts are

$$\Re[\chi_D^{-1}(T)] = \frac{g_0}{1+\omega_0^2\tau^2} \qquad \Im[\chi_D^{-1}(T)] = \frac{g_0\omega_0\tau}{1+\omega_0^2\tau^2} \quad (2A, 2B).$$

at $\omega = \omega_0$. Thus, when one susceptibility component changes due to the $\tau(T)$ term, the other must always change in a quantitatively related manner. Such changes are measurable because

$$\frac{2(f_0 - f(T))}{f_0} = \frac{1}{I\omega_0^2}\Re[\chi_D^{-1}(T)] \qquad D(T) - D_\infty = \frac{1}{I\omega_0^2}\Im[\chi_D^{-1}(T)] \quad (3A, 3B)$$

within the Debye model with suitable approximations (*31*); $D_\infty \equiv \gamma/I\omega_0$. Moreover a well-defined characteristic temperature $T^*$ for such a susceptibility occurs when $\omega_0\tau(T^*) = 1$; both the $f(T)$ slope and the dissipation $D(T)$ achieve their maxima at $T^*$ (Fig. 1).

In Fig. 4A (left) we show a fit of Eq. 3B to the measured $D(T)$ as a red line while Fig. 4A (right) shows the resulting prediction from Eq. 3A for $f(T)$ as the blue line. Comparison to the measured $f(T)$ (solid blue circles) shows that this Debye susceptibility is inconsistent with the relation between $D(T)$ and $f(T)$. Nevertheless, as the relaxation processes of $D(t,T)$ and $f(t,T)$ are synchronized (Fig. 3), there must be an intimate relation between $\Re[\chi^{-1}(t,T)]$ and $\Im[\chi^{-1}(t,T)]$. To study this relation, one should replot the data from Fig. 3, A and B, in the complex plane with axes defined by $\Im[\chi_D^{-1}]$ and $\Re[\chi_D^{-1}]$ [a Davidson-Cole (D-C) plot – (*31*)]. This is a classic technique in which departures of the data from the Debye model appear as geometric features that can reveal characteristics of the underlying physical mechanism linking $\Re[\chi^{-1}(t,T)]$ and $\Im[\chi^{-1}(t,T)]$.

We therefore plot $\Delta D(T) = D(T) - D_\infty$ versus $\frac{2(f_0 - f(T))}{f_0}$ in Fig. 4B. It reveals that, instantaneously upon warming, the D-C plot is a symmetric elliptical curve, whereas after several thousand seconds the response has evolved into the skewed D-C curve more familiar from studies of the dielectric glass transition (*34*). But the maximum frequency shift expected from the maximum observed dissipation within the Debye susceptibility (vertical dashed lines) is again far too small. Moreover, no temperature equilibration lag between the solid $^4$He sample and the mixing chamber could generate the complex dynamics reported in Fig. 4 because, for any given frequency shift, a wide variety of different dissipations are observed (see Supporting Material).

A simple superfluid transition is inconsistent with all these observations because there should be no synchronized dissipation peak associated with $f(T)$ (Figs. 1 and 4) and no ultraslow dynamics in $f(t,T)$ and $D(t,T)$ (Figs. 2 and 3). Indeed, these phenomena are more reminiscent of the characteristics of a glass transition (*34*). Nevertheless, a simple freeze-out of excitations described by a Debye susceptibility is also quantitatively inconsistent because the dissipation peak is far too small to explain the observed frequency shift (Fig. 4A). Thus, when considered in combination with implications of the blocked annulus experiments (*9,16*), our observations motivate a new hypothesis in which amorphous solid $^4$He is a supersolid, but one whose superfluid phase-stiffness can be controlled by the freeze-out of an ensemble of excitations within the solid.

Within such a model generation of excitations at higher temperatures would suppress superfluid phase stiffness. The complex relaxation dynamics (Figs. 3 and 4) would reveal the excitation freeze-out processes. Further, the anomalously large frequency shifts (Fig. 4) would occur predominantly because of superfluid phase stiffness appearing after excitation freezing. Such a model might also explain the diverse phenomenology of solid $^4$He. For example, the ω dependence of $T^*$ (*13*) would occur because $T^*$ is the temperature for which $\tau(T^*)\omega=1$. The shear modulus stiffening (*21*) would occur because of the freeze-out of motion of these excitations, and $T^*$ would increase with $^3$He concentration (*8,11*) because, with pinning, higher temperatures would be required to achieve the excitation rate $\tau(T^*)\omega_0=1$. Finally, sample preparation effects *(10,12,20)* and different responses from different TO types would occur because the amorphousness allowing these excitations would depend on annealing and TO design.

Independent of these hypotheses, important new features of solid $^4$He are revealed here. We find synchronized ultraslow relaxation dynamics of dissipation $D(T)$ and a component of frequency shift of $f(T)$ in TOs containing amorphous solid $^4$He (Fig. 3). Such phenomena are reminiscent of the glassy freeze-out of an ensemble of excitations and inconsistent with a simple superfluid transition. Nevertheless, although the evolutions of $f(T)$ and $D(T)$ are linked dynamically, the situation is also inconsistent with the simple excitation freezing transition because there is an anomalously large frequency shift (Fig. 4). One possible explanation is that solid $^4$He is not a supersolid and that the appropriate rotational susceptibility model for its transition will be identified eventually. But if superfluidity is the correct interpretation of blocked annulus experiments (*9, 16*), then our results indicate that solid $^4$He supports an exotic supersolid in which the glassy freeze-out at $T^*$ of an unknown excitation within the amorphous solid controls the superfluid phase stiffness. Such a state could be designated a "superglass".

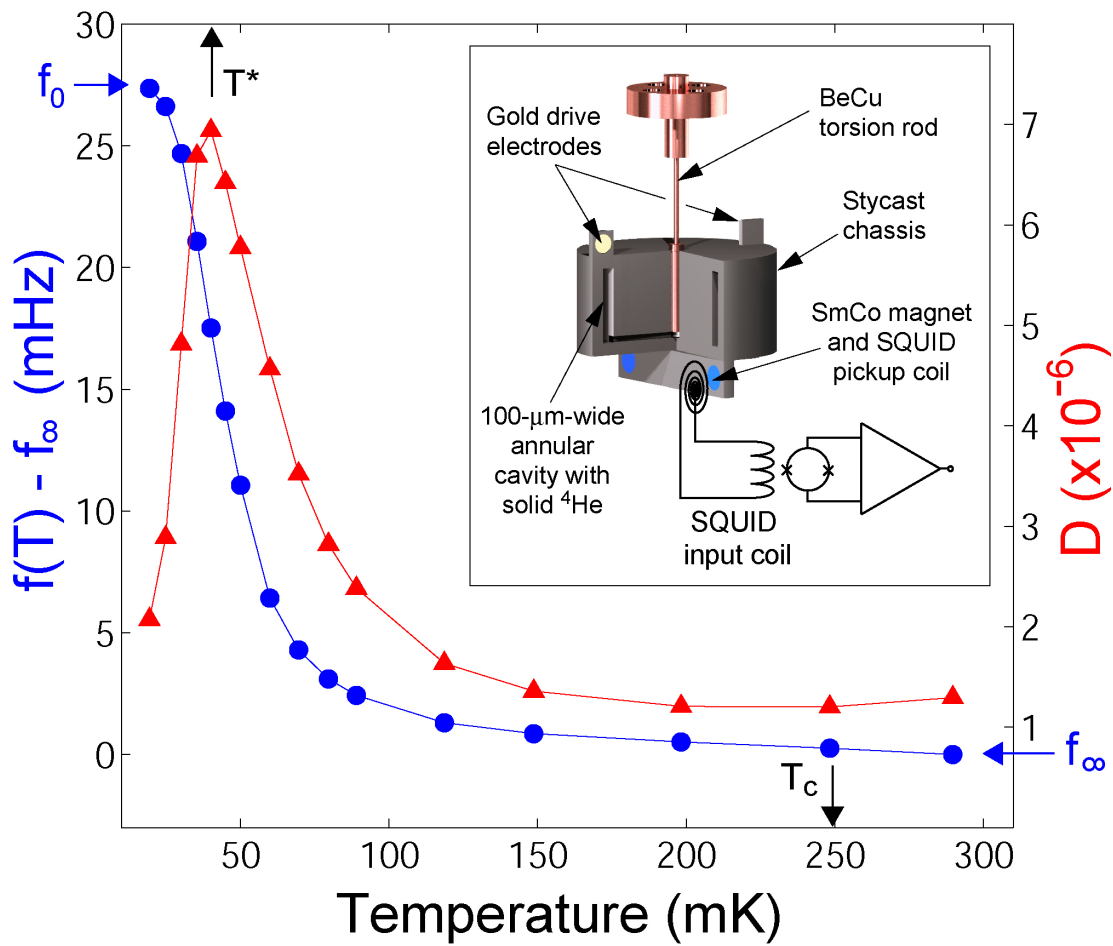

**Fig.1**

The resonant frequency shift $f(T) - f_\infty$ (blue circles) and dissipation $D(T) \equiv Q^{-1}(T)$ (red triangles) for our TO-solid $^4$He system. Indicated with black arrows are $T^*$, the temperature at which $D(T)$ peaks and the slope of $f(T) - f_\infty$ is maximal, and $T_c$, the temperature at which a change in $f(T) - f_\infty$ becomes detectable above the noise. (Inset) A schematic of the superconducting quantum interference device (SQUID) based torsional oscillator (TO). Applying an ac voltage to the drive electrodes rotates the Stycast chassis (containing the solid $^4$He in a 100-μm-wide annular cavity of radius 4.5mm) about the axis of the BeCu torsion rod. The angular displacement of a SmCo magnet mounted on the TO generates a change in the magnetic flux through the stationary pickup and input coils of a dc-SQUID circuit and thereby a voltage proportional to displacement.

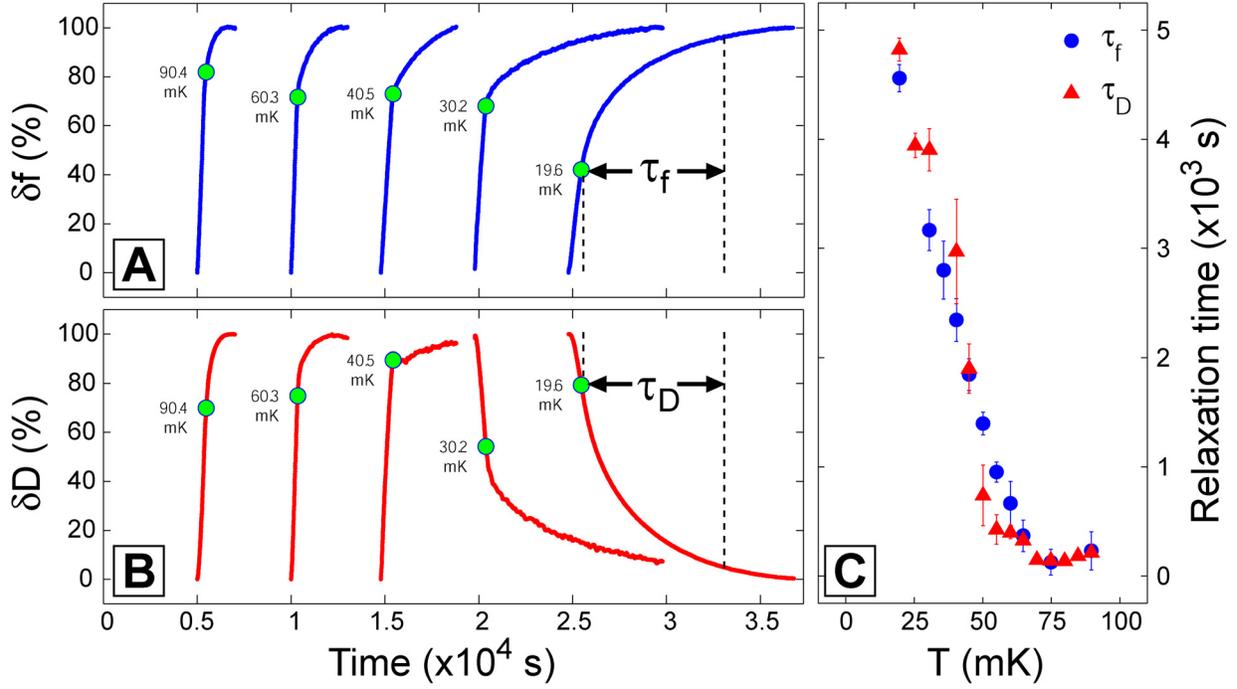

**Fig.2**

Measured traces of **(A)** $f(t,T)$ and of **(B)** $D(t,T)$ for the stepwise-cooling experiment described in the text and fig. S1. These traces are recorded for times $t < t_{eq}$ while the mixing chamber temperature cools from $T_i$ to $T_{eq}$ (before the green dot) and then slow relaxation of $f(t,T_{eq})$ and $D(t,T_{eq})$ for longer times $t > t_{eq}$. The traces are normalized according to $\delta f \equiv [f(t,T) - f(0,T_i)]/[f(\infty,T_{eq}) - f(0,T_i)]$ and $\delta D \equiv [D(t,T) - D(0,T_i)]/[D(\infty,T_{eq}) - D(0,T_i)]$ for comparison at different temperatures $T_{eq}$. **(C)** Measured temperature dependence of $\tau_f(T)$ and $\tau_D(T)$, the relaxation time constants for frequency and for dissipation as defined in the text.

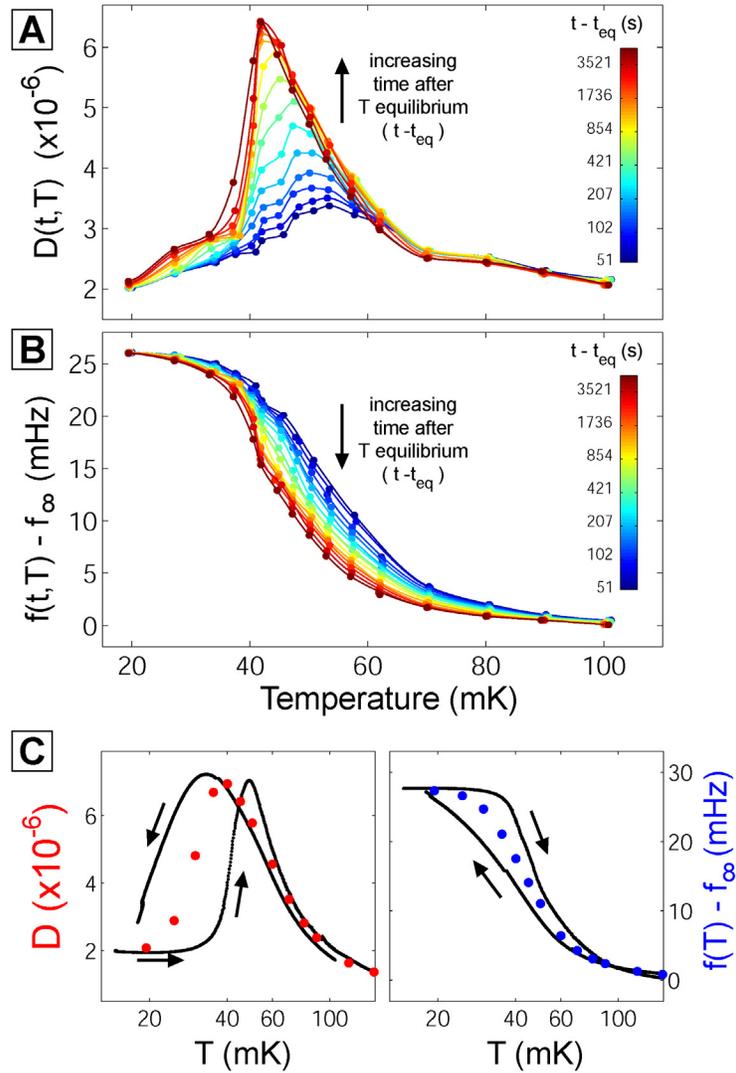

**Fig. 3**

Measured time evolution of **(A)** $D(t,T)$ and **(B)** $f(t,T)$ for the abrupt warming experiment described in the text and fig. S2. The data are colored circles and the lines are smooth interpolations, intended as a guide to the eye. The dark blue lines represent $D(t,T)$ and $f(t,T)$ at $t \sim$ 50s while the dark red lines represent $D(t,T)$ and $f(t,T)$ at ~5000 s. **(C)** Thermal hysteresis in the dynamical response as shown by the black curves in $D(T)$ (left) and in $f(T)$ (right), with the direction of the temperature change indicated by a black arrow. The data indicated by red and blue circles were acquired after waiting t >> $5 \times 10^3$ s at each temperature as the dynamical response [of Fig. 3(A) and (B)] asymptotically approached the infinite-time limit.

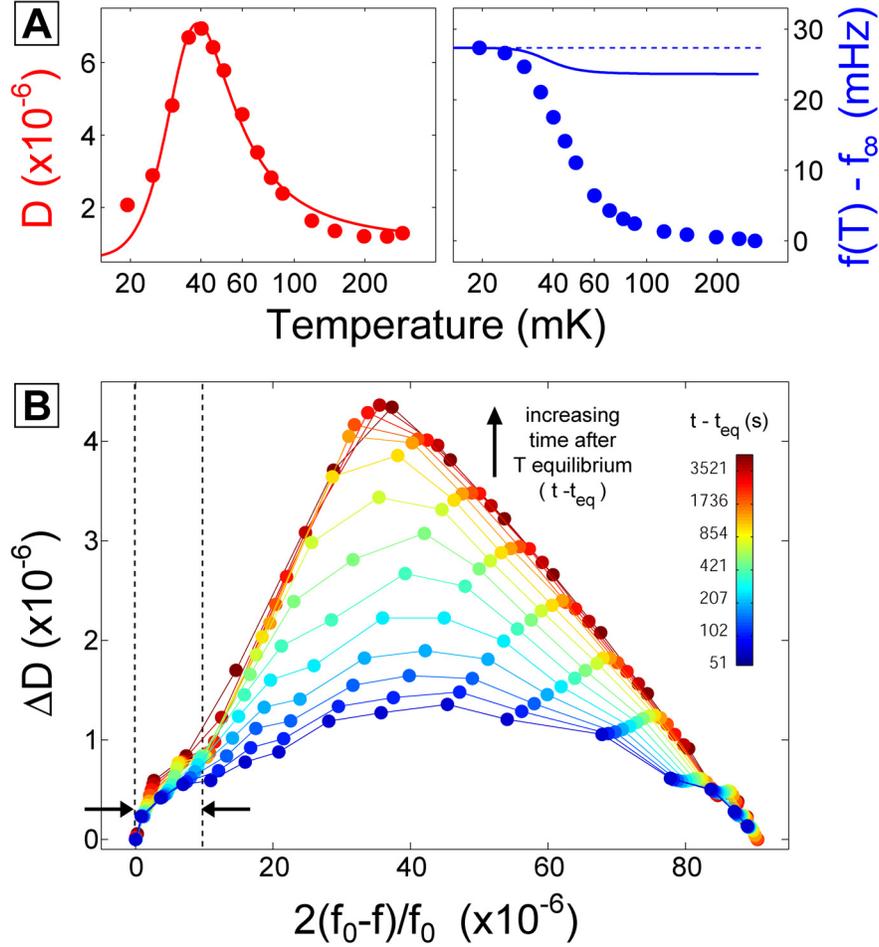

**Fig. 4**

(A) Comparison of equilibrated $D(T)$ and $f(T)$ data with simple Debye model of susceptibility in Eq. 1. The long-time equilibrated data $D(T)$ (left) and $f(T)$ (right) are plotted as filled circles. The dashed line indicates $f_0$. The solid curves represent the predicted relation between the susceptibilities for a simple glassy freeze-out transition. Although the dissipation $D(T)$ can be fit reasonably well by this model (*22*), the magnitude of the frequency shift that is then predicted is markedly smaller than observed. (B) Time-dependent D-C plot. This is a parametric plot of the data shown in Fig. 3, A and B, made by removing the explicit dependence on temperature, with axes defined by $\Im[\chi_D^{-1}]$ and $\Re[\chi_D^{-1}]$ (see Eq. 3). The vertical dashed lines indicate the maximum value of $2(f_0 - f)/f_0$ that would be predicted by the Debye susceptibility (Eq. 3), given the peak height of $\Delta D = D - D_\infty$.

We acknowledge and thank J. Beamish, M. W. H. Chan, A. Clark, A. Dorsey, M. Graf, E. Mueller, S. Nagel, M. Paalanen, R. E. Packard, J. Parpia, J. D. Reppy, A. S. Rittner, J. Saunders, J. P. Sethna, and Wm. Vinen, for helpful discussions and communications. These studies were initiated under National Science Foundation Grant DM-0434801, and are now partially supported under Grant DMR-0806629 and by Cornell University; B.H. acknowledges support by the Natural Sciences and Engineering Research Council of Canada. M.Y. acknowledges support from the Japan Society for the Promotion of Science. Work at Los Alamos was supported by the US Department of Energy.

Supporting Material

# Evidence for a Superglass State in Solid $^4$He

B. Hunt, E. Pratt, V. Gadagkar, M. Yamashita, A. V. Balatsky, and J.C. Davis*

* To whom correspondence should be addressed: jcdavis@ccmr.cornell.edu

## I    Materials and Methods

**(a)    Experimental Procedures**

Our primary observables are the resonant frequency $f$ and the dissipation $D=Q^{-1}$ of a torsional oscillator filled with solid $^4$He. The solid samples are grown from a high-pressure liquid (at ~75 bar and 4.2 K) with a nominal $^3$He concentration of 300 ppb by the blocked capillary method, cooling rapidly along the melting curve (approximately 20 minutes from 4.2 K to < 1 K), and they typically reach a low-temperature pressure of ~36 bar. The samples are formed inside an annular chamber with a cross-section of 0.1 mm x 3 mm and radius of 4.5 mm (see inset Fig. 1), which corresponds to a surface-to-volume ratio of 200 cm$^{-1}$. The torsion rod is made of annealed beryllium copper (BeCu) and the chamber containing the solid helium is made of Stycast 1266. The resonant frequency of the empty cell at 300 mK is 575.018 Hz and that of the full cell at 300 mK is $f_\infty =$ 574.452 Hz. The full-cell $Q$ at 300 mK is $8\times10^5$.

The following two figures (S1 and S2) illustrate the experiments we performed to obtain the results shown in Fig. 2A and B, Fig. 3A and B and Fig. 4B of the main text.

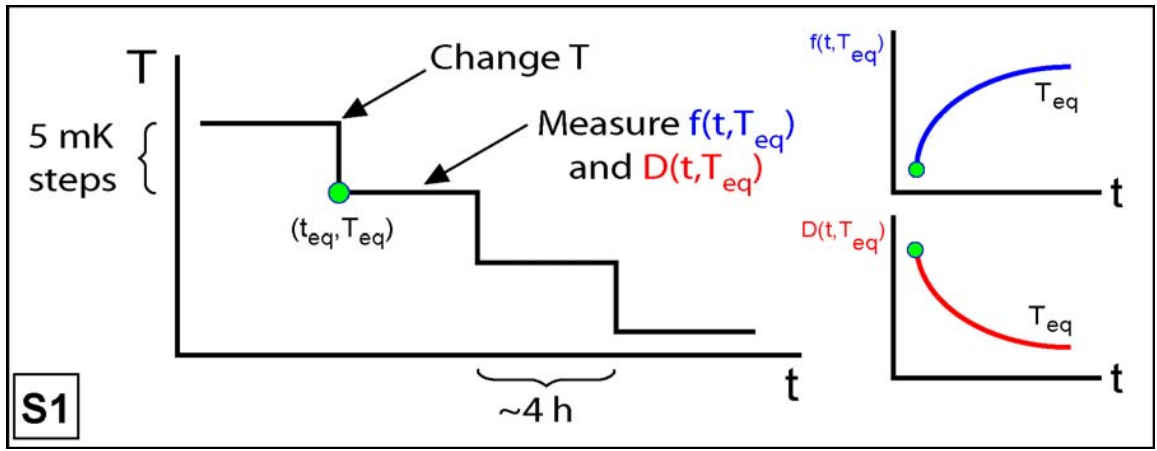

**Figure S1.** The experimental procedure whose results are shown in Fig. 2A and B of the main text. First, the temperature is reduced, typically in a 5 mK step, toward a new target temperature $T_{eq}$. We monitor the frequency and the dissipation during the temperature relaxation for $t < t_{eq}$; during this time the two quantities change at relatively fast rates until the temperature stabilizes at $T_{eq}$. For $T_{eq} < T^*$ (~60 mK), the frequency and dissipation then continue to relax at much slower rates after the temperature has come into equilibrium at the time $t_{eq}$; we record this relaxation $f(t,T_{eq})$ and $D(t,T_{eq})$ for approximately four hours after $t_{eq}$. We find that these phenomena are well fit by an exponential form for the first portion of the relaxation (t<10000 s) and we report the characteristic times $\tau_f(T)$ and $\tau_D(T)$ for these relaxations in Fig. 2C. There are more complex relaxation profiles $f(t,T_{eq})$ and $D(t,T_{eq})$ for much longer times which are not shown in Fig. 2A and B.

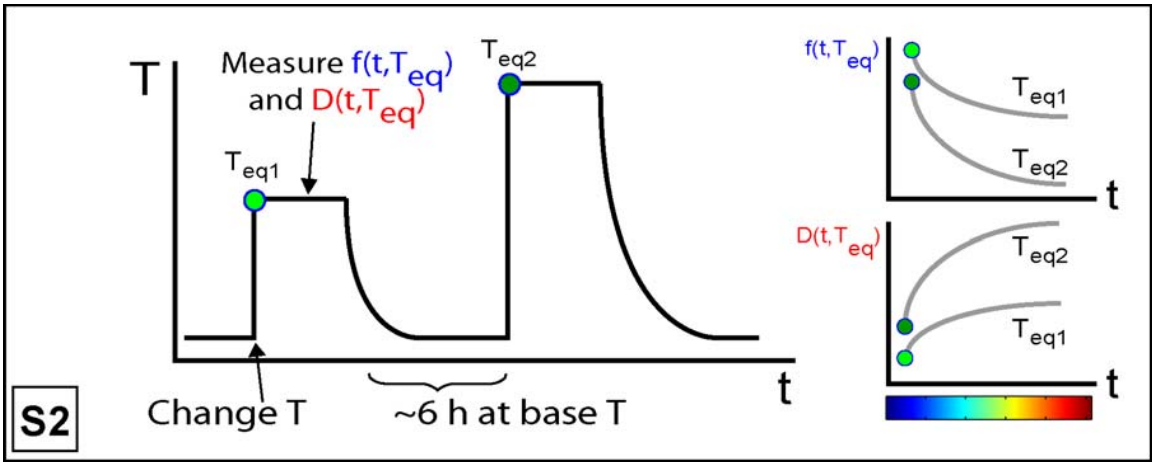

**Figure S2.** The experimental procedure whose results are shown in Fig. 3A and B of the main text. The system is prepared close to its ground state, equilibrating at low velocity and base temperature (about 17 mK) for more than 6 hours. The temperature is then raised rapidly toward $T_{eq1}$ and the frequency and dissipation relaxation $f(t,T_{eq1})$ and $D(t,T_{eq1})$ are recorded for several thousand seconds after the temperature has come into equilibrium. The system is allowed to cool back down to the base temperature and to return to the intial state. The experiment is then repeated for a temperature $T_{eq2} > T_{eq1}$. In this way we are able to create a map of the evolution of the dissipation and the frequency shift due to the solid helium as a function of time after the system is thermally excited from its ground state. The color bar corresponds to the (increasing) times $t_i$ associated with the curves $f(t_i,T)$ and $D(t_i,T)$ in Figure 3A and B, as well as with the Davidson-Cole plots at increasing times in Figure 4B.

**(b)   Linear Response of the Torsional Oscillator - Solid $^4$He System**

The following discussion closely follows that first put forth in Ref. S1 and subsequently in Ref. S3. As many authors have pointed out (S1,S2,S3), what is measured in the torsional oscillator measurements is not the moment of inertia of the TO-helium system but rather its susceptibility $\chi(\omega)$ to an external torque $\Gamma_{ext} = \Gamma_0 \sin(\omega t)$. In the time domain, the angular displacement of the TO is

$$I_{TO}\ddot{\theta}(t) + \gamma\dot{\theta}(t) + K\theta(t) = \Gamma_{ext}(t) + \int g(t-t')\theta(t')dt' \tag{S1}$$

for a linear system invariant under time translation. The second term on the right-hand side of the equation is sometimes called the "back-action" of the helium on the TO chassis: due to its finite shear modulus the helium exerts a moment on the TO. Formulating the dynamical problem in this way allows the entire temperature dependence of the susceptibility to be included in the back-action term $g(t,T)$. Taking the Fourier transform of Eq. S1, we find that $\theta(\omega,T) = \Gamma(\omega)/\chi(\omega,T)$, where

$$\chi^{-1}(\omega,T) = \chi_0^{-1}(\omega) - g(\omega,T) \tag{S2}$$

and $\chi_0^{-1}(\omega) = K - I_{TO}\omega^2 - i\gamma\omega$, the rotational susceptibility of the empty TO, is assumed to be temperature-independent. In this expression, the term describing the helium $g(\omega,T)$ has a temperature-independent part $I_{He}\omega^2$ and a temperature-dependent part $\chi_D^{-1}(\omega,T)$, so that $g(\omega,T) = I_{He}\omega^2 + \chi_D^{-1}(\omega,T)$. The definition of $\chi_D^{-1}$ will be made below. The resonant frequency of the system at the lowest temperature is $\omega_0 = \sqrt{K/I}$, which is that of a perfectly rigid rotor with moment of inertia $I = I_{TO} + I_{He}$. We neglect the small dissipation of the TO $\gamma = I\omega_0 Q_\infty^{-1} = I\omega_0 D_\infty$ in the following because it contributes a correction to the resonant frequency proportional to $D_\infty^2 = Q_\infty^{-2}$, which is $O(10^{-11})$.

The subscript "D" in this particular model of the back-action is because the classic Debye susceptibility is of the form

$$\chi_D^{-1}(\omega,T) = \frac{g_0}{1 - i\omega\tau(T)} \tag{S3}$$

and it describes the freezing out of an ensemble of excitations at a temperature $T^*$ such that $\omega\tau(T^*)=1$. In this case, we find that the real and imaginary parts of Eq. S3 are (for $\omega = \omega_0$)

$$\Re[\chi_D^{-1}(T)] = \frac{g_0}{1+\omega_0^2\tau^2} \qquad \Im[\chi_D^{-1}(T)] = \frac{g_0\omega_0\tau}{1+\omega_0^2\tau^2} \qquad (S4)$$

This is Equation 2 of the main text.

Experimentally, we do not track directly the susceptibility (Eq. S2 and Eq. 1) at the fixed frequency $\omega_0$, but rather we measure the small deviation of the resonant frequency $f(T)$ from its low-temperature, rigid-body value $f_0 = \omega_0/2\pi$, as well as the accompanying peak in the dissipation $D(T)$. Unsurprisingly, perhaps, the quantities $f(T)$ and $D(T)$ are related in a simple way to the real and imaginary parts of the rotational susceptibility within the Debye model, and indeed within any model that responds at a single mode frequency $\tau^{-1}$ (see, e.g., the viscoelastic model of Ref. S3).

The resonant frequency and dissipation are the real and imaginary parts of the complex frequency $\tilde{\omega}$, which is a pole of the function $\chi$ and therefore the solution to the equation $\chi^{-1}(\tilde{\omega}) = 0$. Setting Eq. S2 equal to zero at $\omega = \tilde{\omega}$, we see that it becomes a cubic equation for $\tilde{\omega}$. The exact solution is not illuminating; we can achieve a clearer result by recognizing that $g(\omega,T)$, the contribution to $\chi$ from the solid helium, is a small perturbation to the total susceptibility. We therefore consider a form of $\tilde{\omega}$ that is linearized about the low-temperature solution $\omega_0 = \sqrt{K/I}$,

$$\tilde{\omega} = \omega_0 + ix + y, \quad |ix+y| \ll \omega_0, \qquad (S5)$$

and expand to linear order in $x$ and $y$, giving

$$\tilde{\omega}^2 = \omega_0^2 + 2\omega_0 y + 2i\omega_0 x \qquad (S6)$$

$$\tilde{\omega}^3 = \omega_0^3 + 3\omega_0^2 y + 3i\omega_0^2 x. \qquad (S7)$$

If we substitute these expressions into the cubic equation for $\tilde{\omega}$, we end up with two equations (for $\Re[\chi^{-1}(\tilde{\omega})] = 0$ and $\Im[\chi^{-1}(\tilde{\omega})] = 0$) in two unknowns ($x$ and $y$). Solving for $x$ and $y$ we find that

$$x = -\frac{g_0}{2I\omega_0}\frac{\omega_0\tau}{(1+\omega_0^2\tau^2)} \qquad y = -\frac{g_0}{2I\omega_0}\frac{1}{(1+\omega_0^2\tau^2)} = x/\omega_0\tau \qquad (S8)$$

Using these expressions we can write our observables, the resonant frequency $f$ and the dissipation $D$, as

$$f = \Re(\tilde{\omega})/2\pi \tag{S9}$$
$$= f_0 + y/2\pi$$

and

$$D = |2\Im(\tilde{\omega})/\Re(\tilde{\omega})| \tag{S10}$$
$$\approx D_\infty + 2|x|/\omega_0.$$

In the last line, we added back the contribution to the dissipation from $\gamma = I\omega_0 D_\infty$. It is these expressions that we fit to our data in Fig. 4A, in a slightly different form:

$$f(T) - f_\infty = f_0 - \frac{g_0}{4\pi I \omega_0} \Re\left[\frac{1}{1 - i\omega_0 \tau(T)}\right] - f_\infty \tag{S11}$$

and

$$D(T) = D_\infty + \frac{g_0}{I\omega_0^2} \Im\left[\frac{1}{1 - i\omega_0 \tau(T)}\right] \tag{S12}$$

With the definition of $\chi_D^{-1}$ in Eq. S3, these expressions are equivalent to Eq. 3A and B of the main text. In order to fit the data to Eqs. S11 and S12, we need to assume a specific temperature dependence of the relaxation time $\tau(T)$. In this case, we choose the Arrhenius form $\tau(T) = \tau_0 \exp(\Delta/T)$.

### (c) The Davidson-Cole Plot

We have observed that the (low-velocity) frequency shift and dissipation peak are synchronized and share identical relaxational characteristics as a function of temperature. It seems natural, therefore, to consider whether these observables may be simultaneous consequences of an underlying physical mechanism. In such a scenario, we could expect the temperature to appear as a parameterization between the real and imaginary components of the oscillator's rotational susceptibility. A tool commonly used to illuminate this type of relationship is a direct plot of the imaginary vs. real components of a susceptibility in the complex plane. When used in the context of the dielectric susceptibility $\varepsilon$ of classical glasses or polarized liquids, this is called a Cole-Cole or Davidson-Cole (D-C) plot (S4,S5): the plot is the locus of points $(\Re[\varepsilon], \Im[\varepsilon])$ and the implicit parameter is typically the measurement frequency $\omega$. In our experiment, by contrast, we fix the measurement frequency at the resonant frequency of the system $\omega = \Re(\tilde{\omega})$ and we vary the resonant response of the system at $\tau^{-1}$ by varying the temperature. The analogous plot to $(\Re[\varepsilon], \Im[\varepsilon])$ would be the locus of points $(\Re[\chi^{-1}], \Im[\chi^{-1}])$. The D-C plot is important conceptually because it displays information about the linear response of the system without favoring one implicit variable over another.

As a practical tool, it is also essential. First, in the case of the rotational susceptibility of the TO-helium system, it eliminates the need for specific models of the relaxation time $\tau(T)$. Second, deviations from the Debye susceptibility appear as prominent geometric features in the D-C plot. The Debye susceptibility (Eq. S3) becomes a semicircle centered on $X \equiv \Re[\chi_D^{-1}] = g_0/2$ and with radius $g_0/2$, as can be seen from taking the real and imaginary parts of Eq. S3 and eliminating $\omega_0 \tau$ from the resulting equations. One finds that $(Y \equiv \Im[\chi_D^{-1}])$

$$(X - g_0/2)^2 + Y^2 = (g_0/2)^2. \tag{S13}$$

We know from the previous section and the fits to Eqs. S11 and S12 that the Debye susceptibility is a poor model for our data, but why exactly? To answer this question, we would like to compare our data to the semicircular D-C plot of the Debye susceptibility. Writing Eqs. S11 and S12 in a different form suggests the natural abscissa and ordinate for a Davidson-Cole plot:

$$\frac{2\Delta f}{f_0} = \Re\left[\frac{g_0/I\omega_0^2}{1-i\omega_0\tau}\right] = \Re\left[\frac{\chi_D^{-1}}{I\omega_0^2}\right] \tag{S14}$$

$$\Delta D = \Im\left[\frac{g_0/I\omega_0^2}{1-i\omega_0\tau}\right] = \Im\left[\frac{\chi_D^{-1}}{I\omega_0^2}\right] \tag{S15}$$

where $\Delta D = D(T) - D_\infty$ and $\Delta f = f_0 - f(T)$. This is clear because if the our data were correctly described by the Debye susceptibility, then a plot of $2\Delta f/f_0$ vs. $\Delta D$ would appear as a semicircle. In fact, as anticipated from the failure of Eqs. S11 and S12 simultaneously to fit our data, the situation is quite different.

As mentioned above, departures from the simple Debye susceptibility appear as prominent geometric features in the D-C plot. As one important example, consider a glassy susceptibility in place of the Debye susceptibility, which represents the linear response over a distribution of relaxation times (instead of a single relaxation time $\tau$). Formally, this is obtained by the substitution

$$\frac{1}{1-i\omega_0\tau} \rightarrow \frac{1}{(1-i\omega_0\tau)^\beta} \qquad (S16)$$

with $0 < \beta < 1$ in Eq. S3. Such a susceptibility is manifest as a *skewed* semicircle in the D-C plot (see Fig. S3B below).

We have shown in Fig. 3A and B of the main text that the susceptibility of the TO-solid helium system is time-dependent on extremely long (glassy) time scales. We therefore expect the susceptibility to be representable in a time-dependent D-C plot. Fig. S3A shows the time-independent version of the D-C plot of the glassy susceptibility (16) and Fig. S3B and C illustrate two different possibilities for the ways in which the glassy susceptibility (16) might give a time-dependent D-C plot. Quite simply, there are two parameters, $g_0$ and $\beta$, that can affect the total magnitudes of the real and imaginary parts of the susceptibility and therefore the range, domain and shape of the D-C plot. In Fig. S3B we allow only $\beta$ to vary with time while keeping $g_0$ constant. It is clear that such a model cannot describe our Fig. 4B data accurately; in this model, the height of the D-C plot decreases as time increases whereas in Fig. 4B the height clearly increases with time. The model in Fig. S3C allows $g_0$ to increase and $\beta$ to decrease with time and captures some of the essential features of the time-dependent susceptibility that we measure: the height of the D-C plot increases with time and the skew increases as well. (For clarity, the curves in Fig. S3C are offset so that they are centered at the same point on the X-axis). However, the central question to this work remains: what can account for the disproportionately large extent of the D-C plot along the frequency (X-) direction and why does it not appear to change with time? A naive argument is that most of the frequency shift is due to a superfluid component of the solid helium coexisting with (and whose phase stiffness is apparently controlled by) the glassy component.

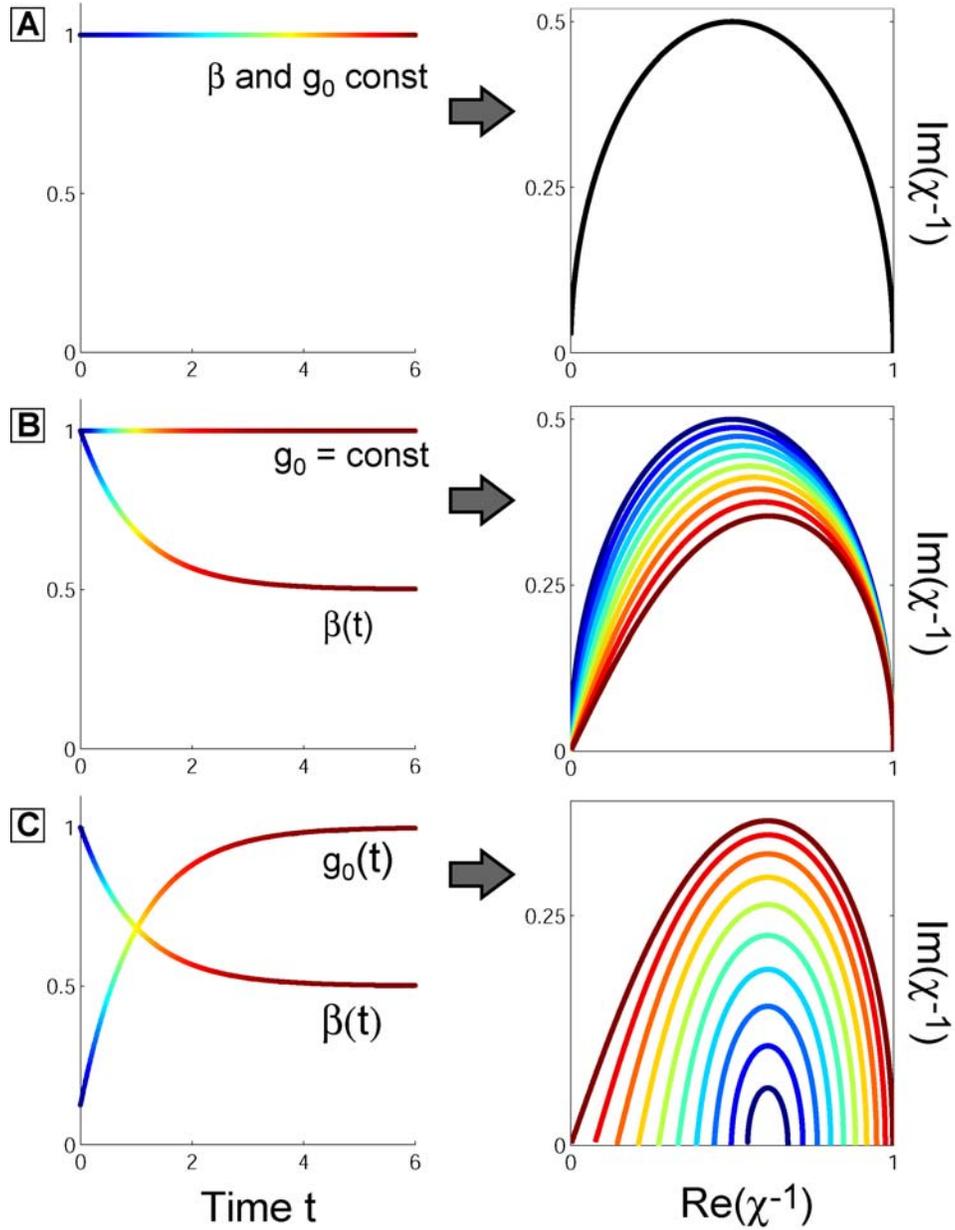

**Figure S3.** Three models of the time dependence of the two parameters, $g_0$ and $\beta$, of a glassy susceptibility (left figures) and their corresponding time-dependent Davidson-Cole (t-d D-C) plots (right figures). The color scale corresponds to the same one as Fig. 4B; blue to red is increasing time. **(A)** For $g_0$ and $\beta$ constant in time, so too is the t-d D-C plot. **(B)** If the glassy exponent $\beta$ is allowed to diminish in time, which describes a spreading out of the modes of the system, the t-d D-C plot skews to the right and its height diminishes as well. **(C)** A model in which both $g_0$ and $\beta$ change as a function of time allows the height of the t-d D-C plot to increase as well as for skew to develop as a function of time, which captures some of the features of Fig. 4B but not all.

## II  Supporting Text

**Thermal Parameters and Time Constants of Experimental Apparatus**

We studied the thermal time constants of the empty torsional oscillator (TO) by gluing a 100Ω Matsushita carbon resistance thermometer onto a test cell with the identical geometry and materials used in the actual TO.  A test cell was used in an effort not to perturb the sensitive parameters of the actual torsional oscillator assembly.  A repeat of the cooling protocol for Experiment 1 (Fig. S1) in this study indicated that there are always long time constants associated with the relaxation of the temperature of the Stycast walls of the empty cell; these are shown in Fig. S4 alongside the full-cell mechanical relaxation data from Fig. 2.  There are several differences.  First, the change in the thermal and mechanical time constants appear to follow different power laws in temperature.  Second, for $T > 75$ mK the thermal time constants are an order of magnitude larger than the mechanical time constants.   Thirdly, and perhaps most importantly, there appears to be no feature in the empty-cell thermal time constant at 75 mK, where the onset of the rise in the full-cell mechanical time constants begins.

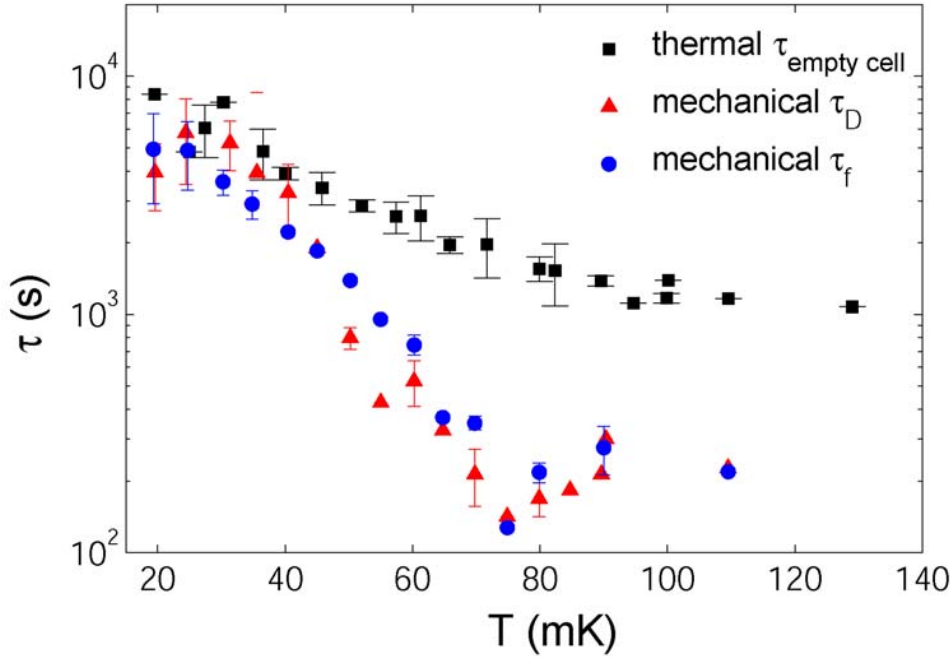

**Figure S4.**  The time constant for thermal relaxation of the empty cell (black squares) compared to the mechanical time constants for the relaxation of the dissipation (red triangles) and frequency (blue circles) of the full cell containing solid helium.

Figure S5 shows the real geometry of the $^4$He-TO system and the location of the materials and thermal quantities of the system.

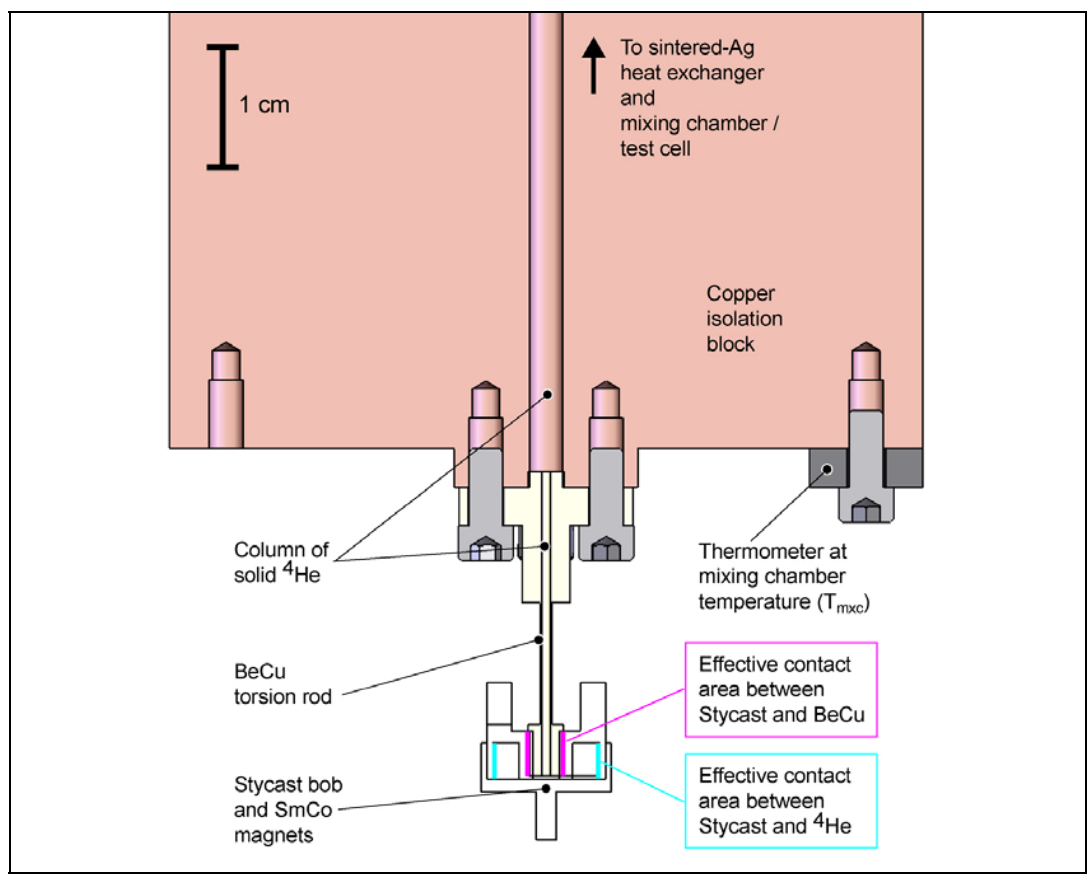

**Figure S5.** Cross-sectional drawing (to scale) of materials in the torsional oscillator - $^4$He apparatus. A correct thermal model of the system would include the temperature-dependence of the heat capacities and thermal conductivities of the copper isolation block, the BeCu torsion rod, the SmCo magnets, the solid $^4$He, and the Stycast 1266 torsion bob, as well as the thermal boundary resistance at the interface between the Stycast and BeCu (shown in cross-section in magenta), the Kapitza resistance between the Stycast and the solid $^4$He (shown in cross section in cyan), and the Kapitza resistance between the BeCu/Cu and the solid $^4$He, whose interfacial area runs the entire length of the torsion rod and the central hole in the isolation block.

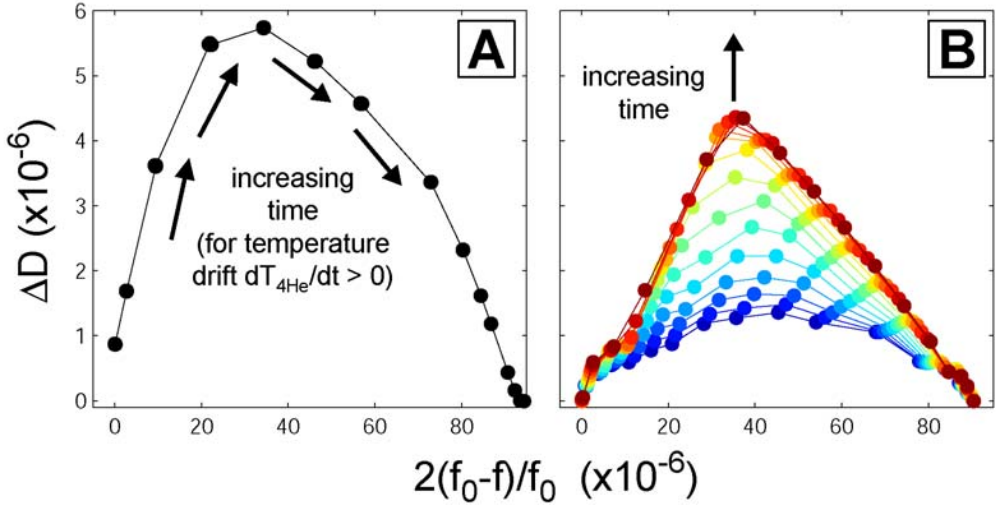

**Figure S6**. **(A)** The time evolution of the Davidson-Cole plot for the dynamics described only by a changing $T_{4He}(t)$. The plot is shown for the case where the temperature increases with time (as would be expected if $T_{4He}$ lagged the mixing chamber temperature $T_{mxc}$ after an increase in $T_{mxc}$, as per the heating protocol of Experiment 2). **(B)** The observed time evolution of the Davidson-Cole plot. These are the same data as Fig. 4B; they reveal a more complicated relationship between the dynamics of the frequency and of the dissipation as the TO-$^4$He system evolves from its low-temperature state.

Since the relative magnitudes of the slow relaxation and fast change in $f$ in Fig. 2A and B indicates that it is indeed the slow relaxation of the solid $^4$He that is contributing predominantly to the relaxation (as opposed to the slow thermalization of the Stycast chassis), there are two possibilities. The first is that the helium has a complicated temperature relaxation function $T_{4He}(t)$, whose precise form depends on the various quantities in the system (Kapitza and thermal boundary resistances, thermal conductivities, heat capacities), and that the dissipation and frequency relaxation data of Fig. 2A and B and of Fig. 3A and B are simply independent measurements of this temperature relaxation function, according to the functionals $D[T_{4He}(t)]$ and $f[T_{4He}(t)]$, whose infinite-time curves would be the results shown in Fig. 1. To test this idea, one would need to measure directly the temperature of the helium within the 100-μm-wide annular cavity, which is at present impossible.

However, there is a simple argument that shows that the complex relaxation dynamics of the solid $^4$He we report (Figs. 3 and 4) cannot be explained by the $^4$He sample being out of thermal equilibrium with the mixing chamber thermometer. For this argument, the time-dependent Davidson-Cole plot (Fig. 4) is an essential tool. The Davidson-Cole plot for the infinite-time curves $D[T_{4He}]$ and $f[T_{4He}]$ is shown in Fig. S6A. One usually thinks of this curve as being parameterized by the temperature of the helium, but if its temperature is changing as a function of time (and $T_{4He}(t)$ is changing slowly compared to $Q/\omega_0$) one can regard as an equivalent parameter the time $t$. This means that the time-dependent Davidson-Cole plot would be indistinguishable from the static (infinite-time) plot – as a function of time the system would simply be moving along the single curve (depicted with a sequence of arrows in Fig. S6A). The data from Fig. 4B, reproduced in Fig. S6B, demonstrate that this is not the case and lead us to the second conclusion: far richer relaxation dynamics exists in solid $^4$He than would be produced by a mere delay in thermalization of the sample.

## III    Supporting References